\documentclass{ar2e}
\usepackage{cite}
\usepackage{mhchem}
\usepackage{graphicx}
\usepackage{caption}

\begin{document}

\input epsf.tex    
\input epsf.def   

\input psfig.sty

\jname{Ann. Rev. Phys. Chem.}
\jyear{2013}
\jvol{XX}
\ARinfo{1056-8700/97/0610-00}

\title{Simulation and Theory of Ions at Atmospherically Relevant Aqueous Liquid-Air Interfaces}

\markboth{}{Atmospherically Relevant Aqueous Interfaces}

\author{Douglas J. Tobias,$^1$ Abraham C. Stern,$^1$ Marcel D. Baer,$^2$ Yan Levin,$^3$ and Christopher J. Mundy$^2$
\affiliation{$^1$Department of Chemistry, University of California, Irvine, California 92697-2025; \newline
$^2$Chemical and Materials Science Division, Pacific Northwest National Laboratory, Richland, Washington 99352; \newline
$^3$Insituto de F{\'i}sica, Universidade Federal do Rio Grande do Sul, Caixa Postal 15051, CEP 91501-970, Porto Alegre, RS, Brazil; \newline
e-mail: dtobias@uci.edu, acstern@uci.edu,marcel.baer@pnnl.gov, levin@if.ufrgs.br, chris.mundy@pnnl.gov}}


\begin{abstract}
Chemistry occurring at or near the surfaces of aqueous droplets and thin films in the atmosphere influences air
quality and climate.  Molecular dynamics simulations are becoming increasingly useful for gaining atomic-scale
insight into the structure and reactivity of aqueous interfaces in the atmosphere. Here we review
simulation studies of atmospherically relevant aqueous liquid-air interfaces, with an emphasis
on ions that play important roles in the chemistry of atmospheric aerosols.  In addition to surveying results
from simulation studies, we discuss challenges to the refinement and experimental validation of the methodology
for simulating ion adsorption to the air-water interface, and recent advances in elucidating the driving
forces for adsorption. We also review the recent development of a dielectric continuum theory that is capable
of reproducing simulation and experimental data on ion behavior at aqueous interfaces.
\end{abstract}

\maketitle

\section{INTRODUCTION}

Aqueous liquid-air interfaces are ubiquitous in the atmosphere.  Common examples include
the surfaces of cloud droplets, aqueous aerosols, thin films of water on boundary layer surfaces
(soils, plant, buildings, etc.), and the quasi-liquid layer on ice and snow.  These aqueous environments
contain myriad chemical species dissolved within them or adsorbed on their surfaces, such as
inorganic atomic and molecular ions, acids and bases, and a wide variety of organic compounds.
Chemical reactions occurring in these environments are involved in both the formation and
removal of trace gases in the atmosphere, and hence contribute to the determination of air quality.
The chemistry of aerosol particles modifies their optical properties and potency as cloud condensation
nuclei, and hence is relevant to radiative forcing and climate.

During the last decade and a half, it has become increasingly apparent that many important
atmospheric chemical processes occur {\em at} aqueous-air interfaces
\cite{kolb95,knipping2000,hunt2004,clifford2007,wingen2008,richards2011}.
Examples of such ``heterogeneous'' chemical processes include the hydrolysis of SO$_2$ in cloud droplets,
which is involved in the production of acid rain \cite{calvert1985}, the oxidation of Cl$^-$ to Cl$_2$ in sea salt aerosol,
which contributes to ozone formation in the polluted marine boundary layer \cite{knipping2000,knipping2003},
and the destruction of tropospheric ozone via halogen chemistry in the Arctic \cite{shepson2007acp}.  It is also
becoming increasingly clear that the aqueous  liquid-air interface provides a unique setting that enables chemistry
that does not occur or is too slow to be relevant in bulk solution \cite{knipping2000,bjfp2009pccp}.

Experimental determination of the kinetics and mechanisms of interfacial chemical processes
is particularly challenging because of the inherent difficulty of making measurements that provide a
molecular-scale view of the composition, structure, dynamics, and chemistry at liquid interfaces.
Consequently, much of the recent progress in the molecular-scale understanding of heterogeneous atmospheric
chemistry has come from molecular dynamics (MD) simulation studies, often carried out in close collaboration
with laboratory studies. In addition to providing insight into particular chemical processes relevant to atmospheric
chemistry, MD simulations have also contributed to the development of fundamental theoretical concepts
that are broadly applicable to a wide variety of problems in interfacial chemistry, and are guiding the
refinement of theories of aqueous solutions, such as dielectric continuum theory (DCT).

In this review, we present and discuss recent simulation studies and developments in dielectric
continuum theory that are relevant to the heterogeneous atmospheric chemistry of aqueous ionic solutions
in the environment.  This is a large and rapidly growing area of research that cannot be covered completely
here.  We have chosen, therefore, to restrict our focus to the behavior of inorganic ions, with an emphasis on
halide anions, at the air-water interface. For a more complete picture, we refer the reader to other recent reviews
that present complementary perspectives \cite{djtreview,bjfp2009pccp,gerber2009irpc,netz2012arpc}.

\section{HALIDE IONS: CONSENSUS VIEW FROM SIMULATION AND EXPERIMENT}

\begin{figure}
  \centering
  \includegraphics[width=4.0in]{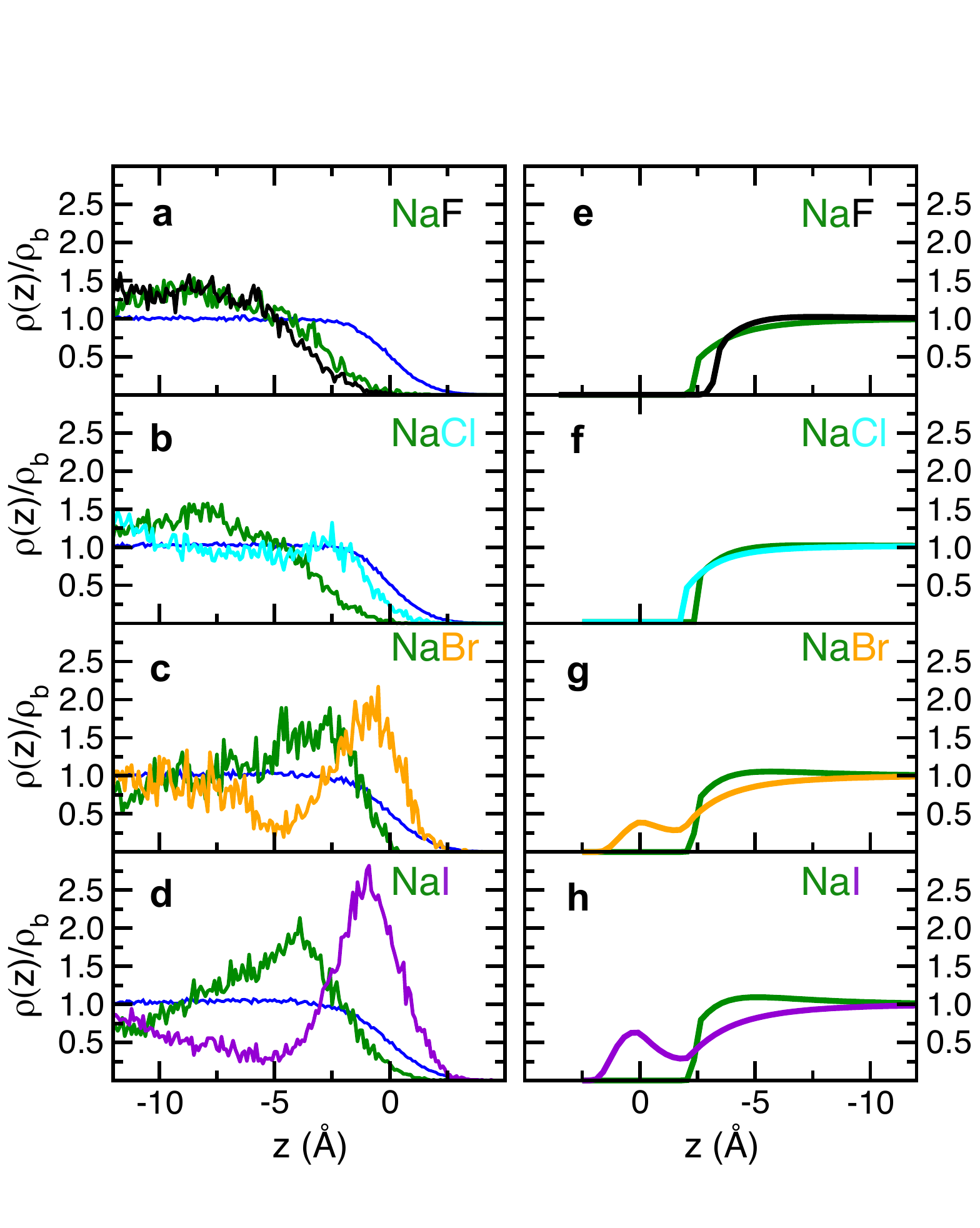}
\caption{Ion density profiles in 1 M sodium halide solutions
 from MD simulations with polarizable force fields (a-d) \cite{djtnewview} and polarizable anion dielectric continuum theory
 (e-h) \cite{LeSa09}. In (a-d) the water density profiles are drawn in blue. The density profiles are normalized by the bulk densities
 and plotted as a function of the
 distance from the air-water interface, defined as the Gibbs dividing surface, located at $z$ = 0. The MD simulation and dielectric   
 continuum theory provide the same qualitative picture of anion adsorption to the air-water interface, namely that large polarizable anions
 can be present at the air-water interface, but the extent of anion adsorption predicted by the polarizable ion dielectric continuum theory
 is substantially less than that predicted by the MD simulations.}
  \label{fig:densityprofs}
\end{figure}

Marine aerosols formed by bursting bubbles in seawater contain halide anions (Cl$^-$, Br$^-$, and I$^-$),
which participate in numerous reactions with atmospheric trace gases, and it is now clear that some
of the rich halide chemistry occurs on the surface of sea salt aerosol particles or snow on which
sea salt has been deposited \cite{bjfp2009pccp,bjfp2010analchem,shepson2007acp}.
A prerequisite for an interfacial reaction is that the reactants (i.e., ions) must be present at the interface.
Until about 20 years ago, this seemed paradoxical, because the conventional wisdom, discussed in more
detail below, was that ions are repelled from the air-water interface.
However, in the early 1990s MD simulations employing polarizable force fields predicted
that the heavier halides prefer to be located on the surface of water clusters \cite{perera,perera1992,perera1993,
caldwell1990,dang1993,stuart1996,stuart1999}, and in the early 2000s
simulations of alkali halide solutions with extended interfaces, modeled as slabs subjected to periodic
boundary conditions, and employing similar force fields, predicted that the heavier halides can adsorb
to the solution-air interface, with an adsorption propensity that follows the order Cl$^-$ $<$ Br$^-$ $<$ I$^-$
\cite{djtnewview,djt2002,dang2002,dang2002a} (Figure \ref{fig:densityprofs}).
These simulations inspired many new experimental investigations, the earliest of which included
second harmonic generation (SHG) \cite{saykally2004,saykally2004a,saykally2006}, vibrational sum-frequency
generation (VSFG) \cite{liu2004,raymond2004}, and photoelectron spectroscopic (PES) \cite{winter2004,ghosal2005}
measurements, that essentially validated the predictions of the simulations.
For reviews of the early simulations of halide ion adsorption and their experimental validation,
see \cite{djtreview,allen2006chemrev,winter2006chemrev,dang2006chemrev,petersen2006}.

\section{MODEL DEPENDENCE OF SIMULATION RESULTS}

During the last decade, there has been explosive growth in the number of simulation studies of halide ion
adsorption to the air-water interface.  The majority of these studies fall roughly into two categories: (1) those
that focus on intrinsic adsorption propensities via calculation of the potential of mean force (PMF) for single
ion adsorption in the absence of counterions; (2) simulations of halide salt solutions at finite
concentration.  The growing body of recent work has exposed a significant level of model dependence of
halide ion adsorption propensity.  Nonetheless, there is general agreement on the order of halide adsorption
propensity, that alkali cations and fluoride anions are repelled from the interface, and that iodide anions
adsorb, although, as we will discuss below, the extent of the adsorption of the heavier halides (\ce{Br-}
and \ce{I-}) is model dependent.

The empirical potential energy functions or ``force fields'' (FFs) that describe interatomic interactions in aqueous
ionic solutions all contain terms that describe hard core repulsive and attractive dispersion interactions, typically
modeled using Lennard-Jones potentials, and Coulomb interactions between fixed charges on the ions and water
molecules.  Ion electronic polarization is typically accounted for explicitly by induced dipoles
\cite{caldwell1990,perera,perera1992,perera1993,dang1993,dang2002,grossfield2003} or Drude oscillators
\cite{stuart1996,stuart1999,roux,warren2008}, in which the atomic polarizabilities are key parameters.
A variety of target data have been used to optimize force field parameters for ions, including quantum chemical
calculations on ion-water complexes, and structural and thermodynamic properties of bulk
electrolytes \cite{dang2002,grossfield2003,horinek2009a}.
Water models are generally not optimized for ion-water interactions and, hence, the resulting ion parameters are tied
closely to the water model with which they are meant to be used.
As has been emphasized in the recent literature, the parameterization of the halide anions is underdetermined,
in the sense that there are degenerate sets of parameters that reproduce available target data equally well
\cite{horinek2009,horinek2009a,netz2012arpc}.
Interfacial properties are not, in general, considered as target data. It is not surprising, therefore, that there is a wide
range of variability in the extent of halide adsorption predicted by different models.

Ion adsorption propensity, quantified in terms of single ion PMFs, varies substantially among the different FFs
that can be found in the literature. PMFs that display minima near the Gibbs dividing surface (GDS) imply that ion adsorption is
favorable thermodynamically, and that the density or concentration of the ion is greater at the interface than
in the bulk.  Non-polarizable FFs typically predict that, among the halides, only iodide is stabilized at the interface
(by $\approx$0.5 kcal/mol), and that the lighter halides, while not stabilized at the interface are able to approach it
more closely in the order \ce{F-} $<$ \ce{Cl-} $<$ \ce{Br-} (see, e.g., \cite{horinek2009}).  Polarizable FFs tend to predict
the same order of closeness of approach, and that, in addition to \ce{I-},  \ce{Br-} is also stabilized at the interface (see, e.g., 
\cite{dang2002}, where single ion PMFs show that \ce{Br-} is stabilized by $\approx$1kcal/mol and \ce{I-} by
$\approx$1.5 kcal/mol).

While systematic variation of force field parameters, such as charge, radius, polarizability useful for
elucidating interactions that contribute to ion adsorption \cite{djtcolloid,geissler1,eggimann2008,horinek2009},
for making predictions that are useful for understanding heterogeneous atmospheric chemistry, accuracy,
as judged by direct comparison to experimental data, is desirable. How well do the disparate halide force
fields perform?  The answer to this question depends, unfortunately, on the data used for the comparison. 

\section{ISSUES CONCERNING EXPERIMENTAL VALIDATION}

The surface tensions of inorganic salt solutions increase with concentration \cite{heydweiller,crittables}.
Surface tension data for halide salts are available over a wide range of concentration, but surface tensions
derived from MD simulations are difficult to converge \cite{dossantos2008,dauria2009}.  Dos Santos et al.
compared surface tensions from MD simulations based on two non-polarizable potentials to experimental data
for NaF and NaI, and noted only mediocre agreement \cite{dossantos2008}.  For NaI at low concentration,
they found that the surface tension was lower than that of neat water, in stark contrast to experimental data.
D'Auria and Tobias computed surface tensions for 1 m and 6 m KF and NaCl solutions using both both
polarizable and non-polarizable FFs and found a similar, mediocre level of agreement with experimental data for
both \cite{dauria2009}. Netz et al. used single ion adsorption PMFs in extended Poisson-Boltzmann (PB)
calculations of ion concentration profiles, from which surface tensions were derived \cite{horinek2007,horinek2009}.
Generally speaking, they found that non-polarizable FFs optimized for solution thermodynamic data
led to better agreement with experimental surface tension data than a polarizable FF.
When comparing PB calculations to experimental data it is worth noting that the PB calculations
appear to underestimate the extent of ion adsorption (and, hence, are expected to overestimate the surface
tensions) compared to MD simulations for a given force field, at least for 0.85 M NaBr and NaI \cite{horinek2009}.

VSFG is a surface-sensitive, nonlinear vibrational spectroscopic technique that has been used extensively
to probe solute-induced changes in water hydrogen-bonding at aqueous interfaces \cite{allen2006chemrev,richmondrev,shen2006sfg}.
VSFG spectra of NaF and NaCl solutions are very similar to the spectrum of neat water, suggesting that
\ce{F-} and \ce{Cl-} do not perturb the water surface \cite{liu2004,raymond2004}.  On the other hand, the spectra of
NaBr and NaI solutions exhibit differences from the neat water spectrum \cite{liu2004,raymond2004,shen2007},
which have been attributed to the presence of the heavier halides in the interfacial region \cite{liu2004,shen2007}.
Accurate calculation of VSFG spectra from MD trajectories is challenging \cite{morita2002,brown2005,buch2007,vanhoucke2009,skinner2009},
but a reasonable level of agreement with experimental VSFG data has been achieved for NaCl and NaI solutions using a
polarizable model that predicted strong adsorption of \ce{I-} \cite{morita2006,ishiyama2007a}.

SHG is a second-order spectroscopic technique that probes electronic transitions in interfacial settings.
SHG has been used to probe bromide and iodide ions at aqueous interfaces \cite{petersen2006,saykally2006,onorato2010}.
To quantify ion adsorption propensities, the concentration dependence of the SHG response is fit to a Langmuir
adsorption isotherm, which affords an adsorption free energy, albeit typically with considerable uncertainty.
Values for \ce{I-} range from $-$6.2 kcal/mol at low concentration (0.1 M to 2 M) to $-$0.8 kcal/mol at higher concentration
($>$ 2 M) \cite{petersen2006,saykally2006}. For \ce{Br-}, the best fit to SHG data from 6 mM to 7 M gave an
adsorption free energy of $-$0.3 kcal/mol \cite{onorato2010}. Thus, consistent with MD simulations employing polarizable
FFs, analysis of SHG data suggests that both \ce{Br-} and \ce{I-} adsorb, and that, for \ce{I-} at low concentration,
the adsorption is much more favorable than single ion PMFs suggest.

Recent advancements in PES experiments carried out at synchrotrons have enabled species-selective depth-profiling
of aqueous solutions \cite{hemminger2006,winter2006chemrev}.  Depth-profiling is possible because as the
incident photon energy is increased, electrons  from deeper in the solution have sufficient kinetic energy to make it to the detector.
Thus, photoelectron kinetic energy (PKE) is a measure of the probing depth. Unfortunately, a precise relation between PKE and
depth in spatial units is not available \cite{ottosson2010a}.  Measurements of the anion/cation ratio as a function of PKE have
been reported for highly concentrated alkali halide solutions.  For KF solutions, the ratio is unity over a range of PKE probing
from near the solution surface to the bulk \cite{Brown-2008}. At low PKE, which corresponds to the interfacial region, the ratio
increases from 1.5 in NaCl to 2 in KBr to 4 in KI \cite{ghosal2005,cheng2012}.
Interfacial anion/cation ratios derived from MD simulations with polarizable FFs compare quite favorably with those
derived from the experimental PES data \cite{Brown-2008,krisch2007,ottosson2010a,cheng2012}
MD simulations have proven useful for enhancing the interpretation of PES data.  For example, the anion/cation ratios
obtained for NaCl and RbBr from PES data and MD simulations suggested no difference in the ion distributions in both
salt solutions, but the MD simulations with polarizable FFs predicted that the \ce{Cl-} interfacial propensity was greater in NaCl
vs. RbCl, and this prompted measurements of \ce{Cl-}/O distributions that confirmed the predictions of the simulation \cite{cheng2012}.

X-ray reflectivity (XRR) measurements probe electron density inhomogeneities at interfaces, and models are required
to extract electron density profiles of the individual species in a multi-component system.  Density profiles extracted
from XRR data for concentrated chloride and iodide solutions suggested that these ions are depleted from the solution-air
interface \cite{sloutskin2007}, in contrast to predictions from MD simulations. Attempts have been made to reconcile MD simulations
with XRR data via comparison of the structure factors rather than the model-dependent density profiles
\cite{dang2009jcp,dang2009jpcb,dang2012}.  These studies have been mostly inconclusive due to lack of quantitative agreement between
the structure factors obtained experimentally and those computed from the MD trajectories.  However, very recently, using a new
polarizable force field, Dang et al. were able to compute structure factors that were in satisfactory agreement with XRR data
for \ce{SrCl2}, thereby establishing that the strong \ce{Cl-} interfacial enhancement observed in the simulation is not
incompatible with XRR data \cite{dang2012}.

Presently, on the whole, surface tension data tend to favor non-polarizable models of halide ions, while surface-sensitive
spectroscopies tend to favor polarizable models. In order to make progress in the validation of FFs used to model halide
solutions, the little or no adsorption inferred from surface tension data needs to be reconciled with the relatively strong
adsorption implied by most available spectroscopic data.  To this end, advances in methodology for calculating spectroscopic
observables from MD simulations and more careful comparisons of simulations to spectroscopic data will be required.

\section{POLARIZABLE ANION DIELECTRIC CONTINUUM THEORY}

Dielectric continuum theories (DCT) for electrolyte solutions have a long and venerable history.
Almost a century ago, Debye and H{\"u}ckel (DH) developed the first simple DCT of bulk
electrolytes \cite{DeHu23}. Within the DH theory, ions are treated as hard spheres with a fixed uniform
surface charge $q$, and water is idealized as a continuum medium of dielectric constant $\epsilon_w$.
DH theory was the first to successfully account for the osmotic properties of bulk electrolytes,
and has served as the starting point for the development of more rigorous statistical mechanical
theories of electrolyte solutions \cite{Le02}.

The development of a successful DCT of the electrolyte-air interface has proven
to be difficult.  The first observation that electrolytes affect the thermodynamics of the
air-water interface has appeared in the pioneering work of Heydweiller, who 
noted that adding a strong electrolyte to water leads to an increase of the  
surface tension of the air-water interface \cite{heydweiller}.  An explanation of this
phenomenon was advanced by Langmuir, who, appealing to the Gibbs adsorption isotherm
suggested that, for reasons that were not very clear, ions preferred
to stay away from the interfacial region \cite{langmuir1917}.
The mechanism responsible for the 
ionic depletion from the interfacial region was elucidated by Wagner \cite{wagner} and 
Onsager and Samaras (OS) \cite{os}.
Using the DCT of DH, these authors suggested that as an ion approaches
the GDS, it is repelled by its image charge located 
on the air side of the interface.
This results in ion-image repulsion, which drives ions away from the interfacial region.
Using DH theory, OS calculated the ion-image repulsion potential assuming zero ionic radius.
Substituting this expression into the Gibbs adsorption isotherm, they were
able to obtain a limiting law for the excess surface tensions of electrolyte solutions,
which agreed quite well with the experimental data for dilute NaCl solutions.  
OS also attempted to extend the range of validity of their expression for surface tension
by including finite ionic radius.  
To this end, they adopted the interionic interaction potential derived by DH for bulk electrolytes.
This improved the value of the calculated surface tension, making it agree with experiments
up to 100 mM. At higher concentrations, however, strong deviations from the experimental data appeared.  

The problem of surface tensions of electrolyte solutions was reexamined by Levin
and Flores-Mena at the turn of the previous century \cite{LeMe01}.  They included two
new ingredients into the OS theory of surface tension: (i) ionic hydration, and (ii) the
polarizability of the counterion screening cloud surrounding each ion. 
Ionic hydration was assumed to prevent ions from coming closer to the Gibbs dividing surface (GDS)
than their hydrated radius, while the broken translational symmetry imposed by the interface was shown to lead to
diminished screening of the ionic self-energy, resulting in an additional repulsion from the GDS. 
The theory proposed by Levin and Flores-Mena was shown to be in excellent agreement with
experimental surface tension data for \ce{NaCl} solutions up to 1M concentrations. 
However, when the same theory was applied to \ce{NaI} solution it predicted a qualitatively
incorrect behavior. Specifically, the surface tension of \ce{NaI} was found to be larger than that of \ce{NaCl},
which was contradicted by experiments.  It was clear that some key ingredient
was still missing from the complete description of the interfacial properties of electrolyte solutions.

A clue to the missing ingredient was provided by the polarizable force field simulations discussed above.  
The prediction that the populations of the large halides (\ce{Br-} and \ce{I-}) are enhanced at the interface vs. the
bulk contradicted the very basis of the OS theory, and polarizability seemed to play a role in promoting
ion adsorption to the interface.
To understand why polarizability is so important to ionic solvation, one needs to compare
the electrostatic self-energy of an ion in vacuum and in bulk water \cite{Le09}.  The self-energy
of an ion of charge $q$ and radius $a$ in vacuum is $U_v \approx q^2/2 a$ while
in water it is $U_w \approx q^2/2 \epsilon_w a$, where $\epsilon_w \approx 80$ is
the dielectric constant of water.  
To move an ion of radius $a\approx 2$~\AA, from water to air, therefore, requires $140 k_B T$ of work.  
This huge energy cost will always favor ionic solvation in bulk water.
Suppose, however, that we want to move an ion not all the way 
into the vapor phase, but only to the interface, so that half of it is still hydrated. The electrostatic
work that is needed to bring a hard non-polarizable ion from bulk electrolyte
to the surface is about $20k_BT$ \cite{TaCo10}.
This is much less than the work needed to move the ion into the vapor, but is still so large that
it makes it energetically improbable to find an ion located near the GDS.   

Now consider a perfectly polarizable ion that can be modeled as a conducting spherical shell of radius $a$
and charge $q$, which is free to distribute itself over the surface of the ion. 
To minimize the electrostatic energy as a polarizable ion moves across the dielectric interface,
its surface charge redistributes, so as to remain mostly hydrated.  The work required to bring a
polarizable ion toward the interface is approximately 10 times smaller ($\approx 2k_B T$)
than that for a non-polarizable ion \cite{Le09,LeSa09}.  Still, from the
point of view of electrostatics, the interfacial ion location is unfavorable.  
What, then, drives large, highly polarizable halogen ions towards the air-water interface?
To answer this question we must consider the energetics of ionic solvation.  To move an ion
from vacuum into bulk water, we must first create a cavity from which water molecules
are excluded, and this is both entropically and energetically costly.  
MD simulations provide an estimate of the cavitational
free energy, which for an ion of $a\approx 2$~\AA, turns out to be about
$2.5 k_B T$ \cite{Le09}. If the ion moves towards the surface, the cavitational free energy
will decrease proportionately to the volume of the ion exposed to air.  For hard,
non-polarizable ions, this is too small to overcome the electrostatic self-energy
penalty of exposing ionic charge to vacuum. 
For soft, polarizable ions, the situation is very different.  For large halogens such as \ce{Br-} and \ce{I-}
the stabilization due to the cavitational energy competes against the electrostatic of self-energy penalty,
since the two are now very similar in magnitude.  Furthermore, the larger the ion the bigger will be the cavitational
incentive for it to move towards the surface and the smaller will be the electrostatic self-energy penalty.
This is the basic idea behind the polarizable anion dielectric continuum theory (PA-DCT) \cite{Le09,LeSa09}.

The general picture that emerges from PA-DCT is that, near the air-water or a general
hydrophobic interface, anions can be divided into two categories: kosmotropes and chaotropes \cite{SaDi10}.
Kosmotropes, such as \ce{F-} and \ce{Cl-} remain hydrated 
and are repelled from the GDS, while large, highly polarizable chaotropes such as  \ce{Br-} and \ce{I-}
shed some of their hydration shell and become stabilized at the interface by the cavitational and
polarization contributions to the free energy (Figure \ref{fig:densityprofs}. The PA-DCT affords the potential of mean
force (PMF) for ion interactions with the interface \cite{Le09}, from which it is possible to obtain ion distributions and,
via integration of the Gibbs adsorption isotherm, the surface tensions of various electrolyte solutions,
which are found to be in excellent agreement with experimental data \cite{LeSa09,SaDi10}.

\section{AB INITIO MOLECULAR DYNAMICS}

{\it Ab initio} or first-principles MD (FPMD) simulations, in which the forces are computed from
the electronic structure, typically determined by density functional theory (DFT), are being
applied increasingly more frequently to study ion behavior in interfacial settings.
Until quite recently, the high computational cost of FPMD simulations limited their
application in this area to ion-water clusters.  For example, a FPMD simulation of
a \ce{Cl-}(\ce{H2O})$_6$ complex confirmed the surface location of the \ce{Cl-}
ion predicted by FF-based simulations, and provided insight into polarization effects
and vibrational spectroscopy of the cluster \cite{djt2001jcp}.  Eight years ago,
Kuo and Mundy pioneered the application of FPMD to an extended aqueous interface,
specifically, an $\approx$35~\AA\ thick slab of 216 water molecules modeled with open
interfaces in a $15 \times 15 \times 70$ \AA\ cell with three-dimensional periodic boundary
conditions \cite{Kuo-2004}.  In addition to affording unprecedented detail into the structure,
hydrogen-bonding, and electronic structure of the air-water interface, Kuo and Mundy's
simulation defined the protocols required to maintain a stable interface with a relatively
small system and to obtain converged results for hydrogen-bond populations and dynamics
\cite{Mundy-2006,kuo2006}.

FPMD simulations hold great potential to elucidate complex
chemistry that can occur in the vicinity of aqueous interfaces, but this potential can
only be realized if the underlying electronic structure method is sufficiently accurate.
Kuo and Mundy's simulation of the air-water interface exposed additional weaknesses
in a generalized gradient approximation (GGA) exchange-correlation functional, specifically,
the BLYP functional, that was commonly used for aqueous systems.  One problem that was
immediately evident was that the equilibrium density of liquid water at ambient conditions  was
$\approx$0.8 g/cc, i.e., significantly less than the experimental value of $1$ g/cc.
This is a serious shortcoming because, without a proper bulk reference state, potentials of
mean force quantifying interfacial propensity are questionable.

Nevertheless, the simulation protocols established by Kuo and Mundy have been used in
studies of the free energetics of adsorption of \ce{F-} and \ce{ClO4-} anions at the air-water
interface. Despite the issue with the bulk density, excellent qualitative agreement with experimentally
inferred interfacial propensities, which follow the Hofmeister series \cite{pegram2007jpcb},
was obtained. Specifically, \ce{F-} was  found to be strongly repelled from the interface while
\ce{ClO4-} was found to have a high interfacial propensity \cite{Brown-2008,Baer-2009}. For
a hydrated ion such as \ce{F-}, it is likely that the bulk density will not qualitatively affect
the strong repulsion that an ion will feel at the air-water due to the fact that the tight
solvation structure is preserved even at a reduced density.\cite{ho2009}  However, problems were
reported in the \ce{ClO4-} study due to the coupling of the anion to strong density fluctuations
of the slab that were attributed to the incorrect bulk density.\cite{Baer-2009}  Artifacts due to the incorrect
bulk density have also been noticed in the solvation of \ce{OH-} \cite{baer2011jcp}.  

Recent FPMD studies have sought to fix one well known deficiency, namely, the 
lack of dispersion, and while this area of research remains active, it has already had
a major impact on the protocol for DFT-based simulations of aqueous interfaces.  With the 
inclusion of an empirical correction for dispersion due to Grimme \cite{grimme2006}, researchers
showed that the density of liquid water utilizing GGA functionals is very close to $1$ g/cc
when simulated under constant temperature and pressure conditons (NpT ensemble)
\cite{schmidt2009}.  As expected, these results carried over to extended slabs, where
excellent agreement with the NpT simulations was obtained \cite{kuhne2011,baer2011jcp}.
Dispersion corrections also led to improvement of other properties of liquid water that were
problematic previously, such as diffusion coefficients and radial distribution functions. 
The extent to which other equilibrium properties are affected by the addition of
dispersion is presently an active line of inquiry. Obviously, this simple extension of DFT
cannot be expected correct all of the known problems, but it does provide a more solid platform
for launching new DFT-based simulation studies of complex phenomena at aqueous interfaces.

The use of DFT for the study of bulk aqueous systems is not without controversy.  Issues
such as slow diffusion and over structuring of bulk aqueous systems have led
to some skepticism.  Additional experimental tests of DFT are highly desirable, and a
couple of recent examples have yielded encouraging results. Modern applications of
multi-edge X-ray absorption fine structure (EXAFS) techniques have allowed researchers
to precisely determine local structure through the assignment of multiple scattering paths
\cite{mccarthy1997,dang2006jpcb}. Two recent FPMD studies of \ce{I-} and \ce{IO3-} solvation
have highlighted the importance of an {\it ab initio} representation of the interaction
potential to describe the solvation structure \cite{baerdang2010,Baer-2011}. In the case of \ce{I-},
a head-to-head comparison of a high quality empirical polarizable potential to DFT
revealed measurable differences in the first solvation shell \cite{baerdang2010}.
Although the differences were subtle, they came into full relief when examining the statistics of
hydrogen bonding.  The differences between the two interaction potentials were
traced to a larger induced dipole for the polarizable model (by a factor of two), which was
also noted in an independent investigation \cite{beck}. This additional asymmetry under
bulk solvation may have implications for surface propensity as well \cite{wick2011}.
The study of \ce{IO3-} revealed a surprising and novel solvation structure
where the \ce{I} center takes on a partial positive charge under bulk solvation and
strongly coordinates three water molecules, while the negative partial charges
on the oxygens remain weakly hydrated \cite{Baer-2011a}.  It was proposed that the strongly
hydrated \ce{I} atom is responsible for the surprising classification of \ce{IO3-} as a kosmotrope, in
agreement with the Jones-Dole $B$-coefficient based on viscosity studies of electrolytes.
It would be a challenge to describe the complex solvation of \ce{IO3-} revealed by
DFT accurately with empirical potentials.

\begin{figure}
  \centering
  \includegraphics[width=3.0in]{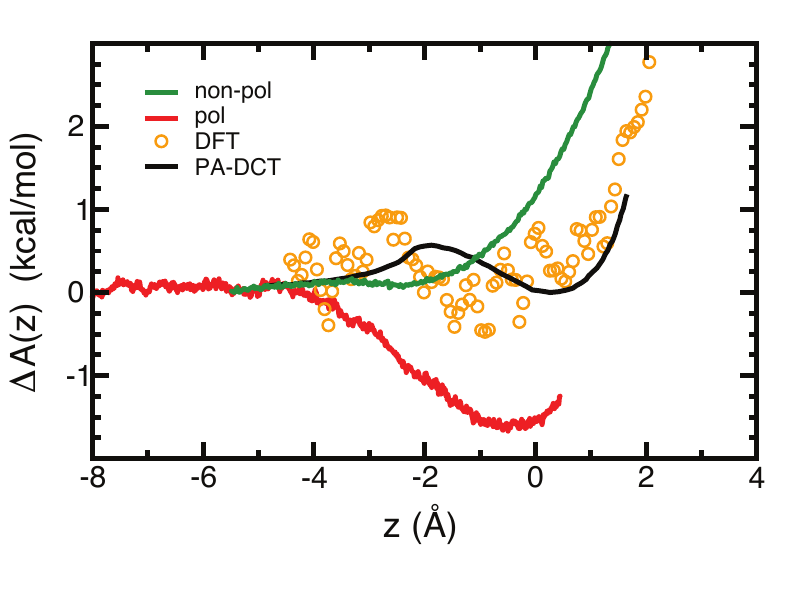}
  \caption{Comparisons of the potentials of mean force for \ce{I-} adsorption at the air-water interface
  (equivalent to the change in Helmholtz free energy accompanying the transfer of the ion from the middle
  of the water slab to a position $z$ in the direction normal to the interface).
  The interface (Gibbs dividing surface) is located at $z$ = 0. The black curve is from the polarizable anion
  dielectric continuum theory \cite{LeSa09}, the yellow curve from density functional theory-based FPMD 
  simulations \cite{Baer-2011}, the red curve from an MD simulation using an empirical, polarizable potential
  \cite{dang2002}, and the green curve from an MD simulations using an empirical, non-polarizable potential
  \cite{horinek2009}.}
  \label{fig:pmfs}
\end{figure}

The favorable comparison with EXAFS data has established that FPMD simulations
with dispersion corrections can reproduce subtle details of the solvent shell accurately
\cite{baerdang2010,Baer-2011a}. The same DFT approach was used recently
to calculate the PMF for \ce{I-} adsorption at the air-water interface \cite{Baer-2011}.
Consistent with early studies that predicted \ce{I-} adsorption using polarizable FFs
\cite{djtnewview,dang2002,dang2002a}, the DFT PMF displays a local minimum at the interface
(Figure \ref{fig:pmfs}). However, the minimum is much shallower than polarizable FFs predict, and actually
resembles PMFs obtained from non-polarizable FFs much more closely \cite{horinek2009}.  Recall that the
latter, when used in extended PB calculations, show reasonable agreement with surface tension data.
The shallow DFT PMF is also strikingly similar to the result from the PA-DCT (Figure \ref{fig:pmfs}), which reproduces
experimental surface tension and surface potential data remarkably well \cite{LeMe01}.

The similarity of the PMFs obtained using DFT, non-polarizable FFs, and PA-DCT is intriguing,
and it raises a number of questions concerning the physics required to describe ion adsorption
correctly. {\em A priori} it is safe to say that the physics contained in DFT is the most complete, but
residual inaccuracies in the functionals and relatively very limited sampling (i.e., compared to
FF-based MD simulations) are causes for concern that will need to be addressed in future studies.
While it is encouraging that the extended PB calculations and the PA-DCT reproduce thermodynamic
data for ion adsorption, it is not obvious why the single ion PMF used in the former, which is derived
from a non-polarizable FF-based MD simulation \cite{horinek2009}, is so similar to the PMF for the
latter, which is obtained via a DCT for polarizable anions \cite{LeMe01}.

\section{DRIVING FORCES FOR ION ADSORPTION}

As evidence continues to mount from both simulation and experiment that certain ions have a propensity
to adsorb at the air-water interface, the discussion has turned away from the question of whether or not
ions adsorb and toward the question of what are the driving forces for ion adsorption. In light of the wealth
of experimental and computational data, it seems that a consensus on the answer to this question
ought to be on the horizon but, to the contrary, recent theoretical studies have come to quite different
conclusions. In the earliest simulation studies that employed polarizable empirical force fields, electronic
polarization and ion size were identified as having an important influence on the propensity for anions
to adsorb to extended aqueous interfaces \cite{djtcolloid}.  Until recently, it was assumed that ion
adsorption is opposed by electrostatics, but that induction effects could reduce the electrostatic
penalty, and favored by hydrophobic/cavitational forces \cite{archontis2006,geissler1}.
More recent studies, reviewed briefly below, have carefully evaluated these contributions and others, including
relative importance of ion-water and water-water interactions, local solvation effects, and the surface potential.
The starting point for most analyses of driving forces for ion adsorption is the single ion PMF, which is either
decomposed into enthalpic and entropic components that are further dissected \cite{herce,ohmine,vanderspoel1,geissler2012},
or partitioned into free energy components \cite{beck2012,baer2012}.

In the first approach, the entropic contribution to the free energy $-T\Delta S(z)$ is determined by subtracting
the enthalpic contribution, which is essentially the average potential energy (assuming the $pV$ term is negligible),
$\langle \Delta U(z)\rangle$, from the PMF, $\Delta A(z)$, where $z$ is a coordinate defining the position of the ion
in the slab, and the $\Delta$ denotes a difference with respect to the bulk reference state. Assuming that the
temperature dependence of the free energy of adsorption can be measured, all of these quantities are
accessible experimentally.  Caleman et al. recently reported a thermodynamic decomposition of single halide
ion PMFs, computed using MD simulations with polarizable FFs in large water clusters, into enthalpic and
entropic contributions \cite{vanderspoel1}. Contrary to previous assumptions, they found that the favorable
adsorption of the heavier halides is favored by enthalpy and opposed by entropy. Otten et al. came to the same
conclusion based on simulations of fractionally charged iodide-like anions \cite{geissler2012}.
They also employed resonant UV second harmonic generation spectroscopy to measure
the adsorption of thiocyanate ions to the interface as a function of temperature \cite{geissler2012}. 
Fitting the SHG response to a Langmuir adsorption model allowed them to obtain changes in free energy,
enthalpy, and entropy for the adsorptive process.  The experimental results confirmed the simulation
results, namely, that the negative free energy of adsorption of the chaotropic \ce{SCN-} anion to the air-water
interface is a consequence of the negative enthalpy of adsorption dominating the negative entropy of adsorption.

\begin{figure}
  \centering
  \includegraphics[width=5.0in]{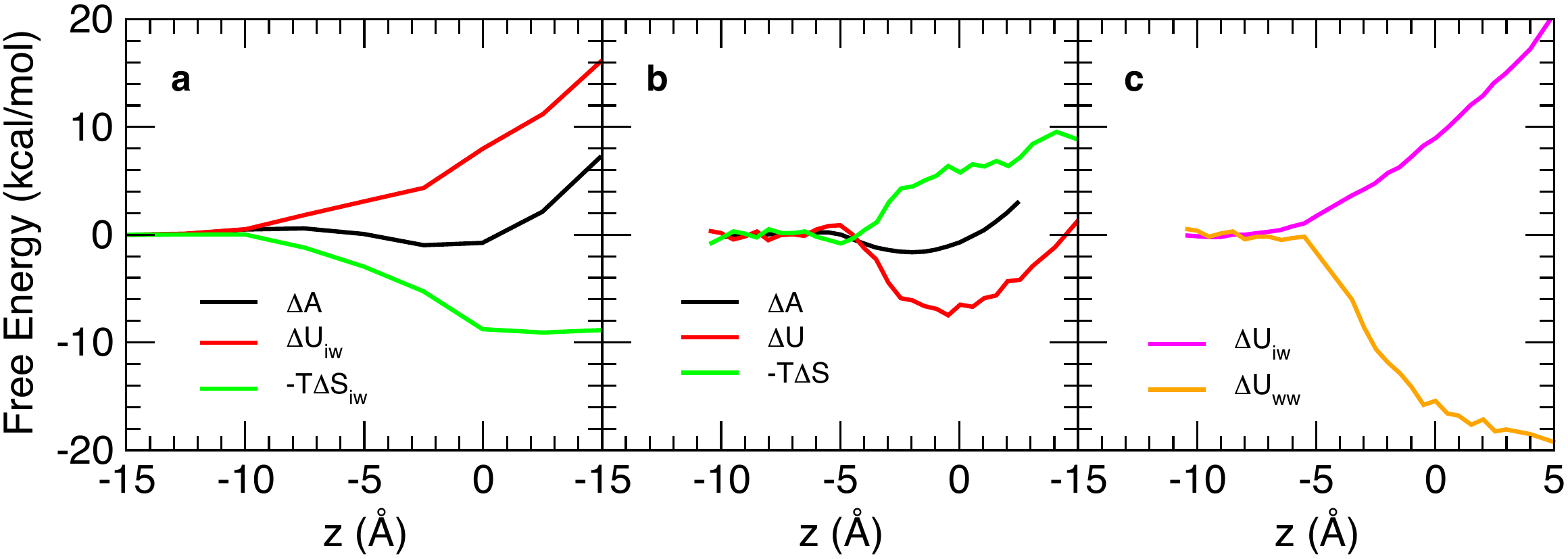}
  \caption{Thermodynamic decompositions of potentials of mean force for \ce{I-} adsorption computed from
  MD simulations with polarizable force fields. The interface (Gibbs dividing surface) is located at $z$ = 0.
  (a) Decomposition into ion-water enthalpy, $\Delta U_{iw}$, and entropy, $\Delta S_{iw}$, by
  Yagasaki et al. \cite{ohmine}. (b) Decomposition into the total enthalpy, $\Delta U$, and total
  entropy, $\Delta S$, by Caleman et al. \cite{vanderspoel1}. (c) Decomposition of the enthalpy into
  ion-water and water-water contributions, $\Delta U_{iw}$ and $\Delta U_{ww}$, respectively,
  by Caleman et al. \cite{vanderspoel1}.}
  \label{fig:decomp}
\end{figure}

The enthalpy and entropy can be further decomposed into contributions from ion-water and water-water
interactions. While the meaning of the individual terms in the enthalpic decomposition is clear, that of
the entropic terms is not. Several authors have asserted that a ``solvent-solvent'' contribution to the solvation
entropy, attributed to the reorganization of the solvent in the presence of a solute, is exactly cancelled by
the solvent-solvent contribution to the enthalpy, i.e., the change in the solvent-solvent interaction energy that
accompanies the insertion of a solute into the solvent (see, e.g., \cite{yukarplus,guillot1993,benamotz2005}).
In this case, the solvation free energy of a single ion in neat water contains only contributions from changes
in the ion-water interaction energy, and an ion-water contribution to the entropy whose physical meaning is not
clear. In other words, changes in water-water interactions do not contribute to the solvation free energy, and hence
do not constitute a driving force for solvation.
In this vein, Yagasaki et al. decomposed single halide ion PMFs, computed from both polarizable and non-polarizable
FF-based simulations and QM/MM simulations in which the ions and their first solvent shell were modeled
using Hartree-Fock theory with a 6-31G* basis set, into ion-water contributions to the enthalpy and entropy,
the latter obtained as the difference between the free energy and the ion-water contribution to the enthalpy \cite{ohmine}.
They found that, for all of the models of all of the ions considered (\ce{F-}, \ce{Cl-}, and \ce{I-}), ion adsorption
is opposed by the ion-water contribution to the enthalpy, as expected, and favored by the contribution to the
PMF from the ion-water entropy (Figure \ref{fig:decomp}). Yagasaki et al. also found that polarization contributes to stabilizing the
interfacial location of the heavier halides, presumably by delaying the enthalpic cost of exposing the charge to vacuum.

A quite different set of conclusions emerges when the water-water contributions to the enthalpy are kept
in the picture. Recall that both Caleman et al. \cite{vanderspoel1} and Otten et al. \cite{geissler2012}
determined that the total enthalpy of adsorption is negative for halide and halide-like anions. Furthermore,
they decomposed the enthalpy into ion-water and water-water contributions, and found that
the positive ion-water contribution (due to desolvation) is more than compensated by a negative
contribution from water-water interactions (Figure \ref{fig:decomp}).  Thus, they singled out changes in water-water
interactions as a driving force for ion adsorption.  Otten et al. proceeded to show that water-water interactions
are less favorable in the first solvation shell of the ion and at the air-water interface than they are
in bulk water, and that, as an anion approaches the interface it sheds some of its solvation shell and
displaces water from the interface into the bulk \cite{geissler2012}.  The negative water-water contribution
to the adsorption enthalpy results, therefore, from a loss of less favorable hydration shell and surface water-water
interactions, and an increase in more favorable water-water interactions. Otten et al. concluded that the
enthalpic driving force for ion adsorption arises from changes in local changes in solvation.

Otten et al. examined two possible explanations for the negative adsorption entropy \cite{geissler2012}.
The first, a change of the orientational statistics of water molecules in the ion hydration
shell, led to an estimate of the adsorption entropy that erred both in sign and magnitude
from the value extracted from the temperature dependence of SHG data.  The second,
a reduction of entropy due to pinning of capillary waves by an ion at the interface, estimated
by applying a harmonic model to the interfacial height fluctuations, led to a value that
had the correct sign and was within a factor of two of the experimental value for the
adsorption entropy.

Rather than focusing on enthalpic and entropic contributions to the free energy,
Arslanargin and Beck used a free energy partitioning approach
to identify the predominant driving forces for anion adsorption \cite{beck2012}.
According to their analysis, the stabilization of a non-polarizable iodide anion at the 
air-water interface results primarily from the reduction of the size of the cavity in water that
accompanies adsorption, and a ``far-field'' electrostatic contribution corresponding
to the ion interaction with the interfacial surface potential.  Around the same time, 
Baer et al. came to a similar conclusion, and showed that anion adsorption free energies
obtained from MD simulations could be accurately reproduced by a DCT
that includes free energy terms for cavity formation, electrostatic self energy, and
a modest electrochemical surface potential ($\approx$$-$0.3 V, which is about half
the value of the surface potential of the water model used in the MD simulations)
\cite{baer2012}.

The electric potential difference across the air-water interface arises
because of the net orientation of water molecules in the interfacial region
\cite{wilson1988,netz2012arpc}.
Surface potentials computed from MD simulations of point-charge water models
tend to be around $-$0.5 V, and the analyses summarized above agree that the
surface potential can be considered as a significant driving force for anion adsorption.
The role of the surface potential becomes less clear, however, when results from FPMD
simulations and PA-DCT are included in the discussion. The PMF for iodide adsorption
derived from PA-DCT does not include a surface potential term, yet it agrees remarkably
well with the iodide PMF computed from FPMD simulations (Figure \ref{fig:pmfs}), and reproduces
experimental surface tension accurately \cite{LeSa09}. This juxtaposition of results
suggests that the electrochemical surface potential felt by an adsorbing ion is negligible.
However, the surface potential calculated explicitly from FPMD simulations is large and positive
\cite{kathmann}, and, therefore, should strongly oppose anion adsorption. This is not what is
observed, e.g., in the PMF for \ce{I-} adsorption \cite{Baer-2011} (Figure \ref{fig:pmfs}), and
reconciling this discrepancy is a subject of current research.

\section{SOLUTIONS CONTAINING MIXTURES OF SALTS}

While much has been learned about the distributions of ions near atmospherically relevant aqueous
interfaces by studying single-component electrolyte solutions, aqueous solutions in the atmosphere
are multi-component systems.  For example, while the most prevalent ions in seawater are
sodium cation and chloride anion, seawater also contains appreciable amounts of sulfate dianions and magnesium
dications \cite{crchandbook}.  Bromide anions, while present in seawater in only trace amounts ($\approx$1/600 the chloride
concentration), are important because they play a major role in the chemistry of sea salt aerosol \cite{bjfp2000jpca,
bjfp2003chemrev}.  Reactions of gaseous nitrogen oxides with halides in sea salt aerosols leaves behind nitrate anions,
which can participate in photochemical reactions \cite{bjfp2009pccp}.  For a complete description of the heterogeneous
chemistry of aqueous ionic solutions in the atmosphere, it is necessary to understand how the distribution of one species is affected
by the presence of others.  This issue has been addressed in several recent MD simulation studies.

Bromide enhancement at the air-water interface has been proposed to explain its high reactivity in sea salt
aerosols and on snow \cite{bjfp2009pccp,bjfp2010analchem,shepson2007acp}. In the first MD simulations
of systems containing more than one polarizable halide, Jungwirth and Tobias \cite{djt2002} found that,
in equimolar mixtures, chloride and bromide behaved as they did in single component solutions at low concentration
(0.3 M), but at higher concentration (3.0 M), the interfacial population of bromide was strongly enhanced,
and chloride ions were pushed into the bulk.  Two subsequent studies employed a combination of
PES and MD simulations to further explore ion surface propensities in solutions
of chloride/bromide mixtures.  Ghosal et al. \cite{ghosal2008} obtained PES data on deliquesced and dissolved
NaCl crystals doped with 7\% or 10\% Br$^-$ and found that, in each case, the Br$^-$:Cl$^-$ ratio was greater at
the solution interface compared to in the parent crystal.  In the case of the crystal containing 7\% Br$^-$,
they were also able to show that the Cl$^-$:Na$^+$ ratio was significantly reduced (by $\approx$50\%)
at the surface of the solution vs. the crystal. The ion distributions derived from a simulation of a mixed NaCl/NaBr
solution containing 10\% Br$^-$, and employing a polarizable force field, agreed qualitatively with the PES data.
The simulation actually under-predicted the surface enhancement of Br$^-$, and the discrepancy was attributed
to differences in the total ion concentrations in the simulation and experiment, as well as uncertainties in
converting the spatial distributions from the simulations to distributions as a function of photoelectron kinetic
energy for direct comparison with the PES data. Ottosson et al. also used PES and MD simulations to
determine ion distributions in solutions containing mixtures of Cl$^-$ and Br$^-$, and came to similar conclusions
as the previous studies, namely, that Br$^-$ is more surface enhanced in the mixtures vs. in a neat solution of
Br$^-$ at a given concentration, i.e., NaCl ``salts out'' Br$^-$ to the surface, and this effect increases with
concentration \cite{ottosson2010}.  The preference of Cl$^-$ for the interior and Br$^-$ for the surface of the
solution was rationalized in terms of the stronger hydration of Cl$^-$ vs. Br$^-$.  A conclusion common to all
the studies of Cl$^-$/Br$^-$ mixtures is that, at high ion concentration, the behavior of individual ions in mixed
salt solutions cannot be extrapolated from their behavior in neat solutions of a single salt.

In light of the recent appreciation of the importance of iodine in halogen chemistry and particle formation in
coastal environments \cite{saiz-lopez2007,huang2010}, Gladich et al. performed a MD simulation study of solutions of
ternary mixtures of NaCl, NaBr, and NaI modeled with a polarizable force field \cite{carignano2011jpca}.  In an equimolar
mixture of the three components, the anion surface propensity followed the expected order: Cl$^-$ $<$ Br$^-$ $<$ I$^-$.
While increasing the concentration of Br$^-$ (keeping the total ion concentration constant) did not result in changes of surface
propensity, increasing the concentration of Cl$^-$ led to strong surface segregation of Br$^-$ and I$^-$.  Thus, consistent with
simulations of Cl$^-$/Br$^-$ mixtures, Cl$^-$ was observed to ``salt out'' the heavier halides in the ternary mixtures.
In a control simulation of the equimolar mixture with a non-polarizable force field, all of the ions behaved the same:
none adsorbed to the interface, but rather all remained well-solvated in the interior of the solution, underscoring,
again, the profound role that polarization plays in driving ion adsorption in MD simulations.

\begin{figure}
  \centering
  \includegraphics[width=4.0in]{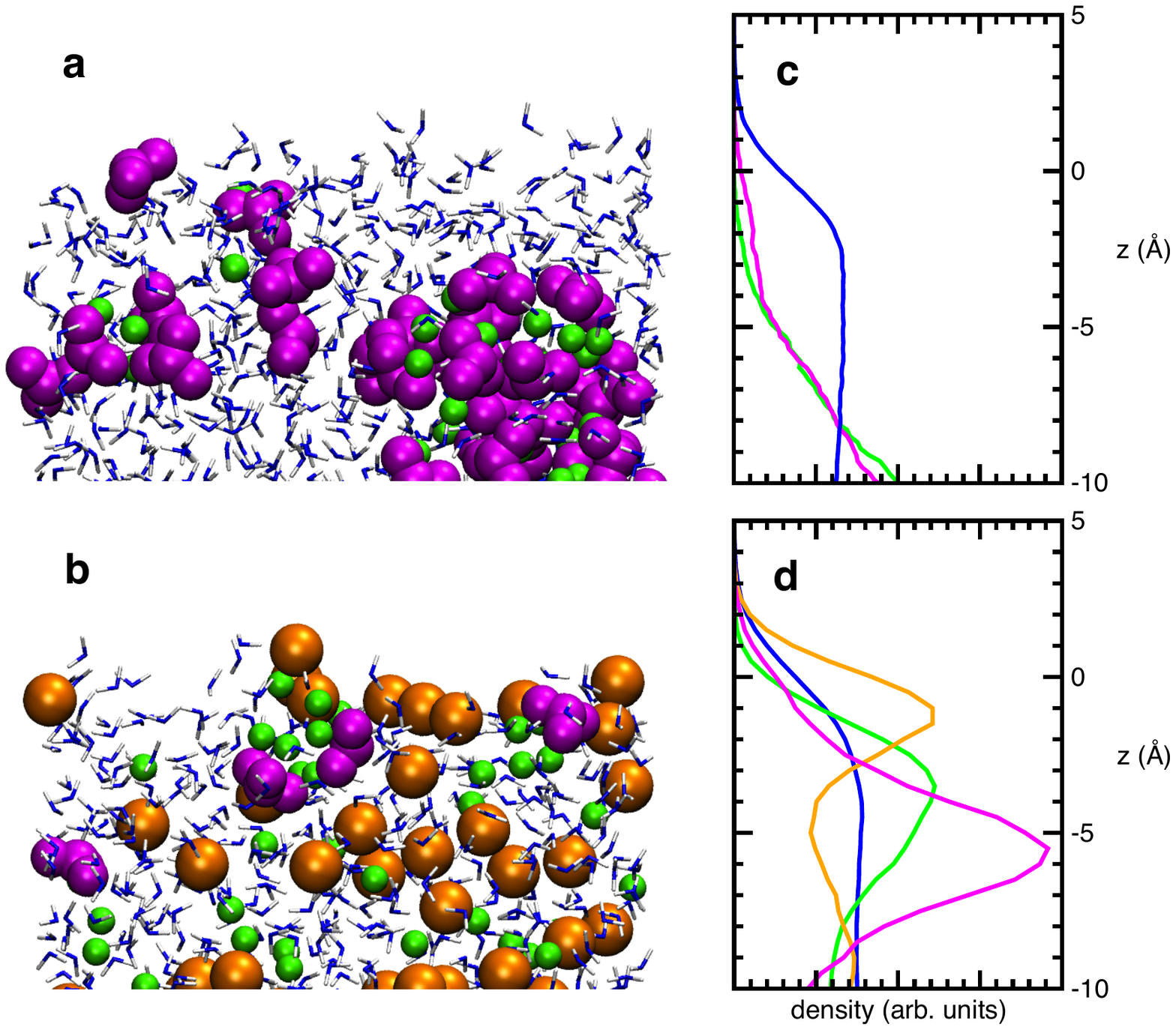}
  \caption{(a) Snapshot and (b) density profiles from a MD simulation of 4 M \ce{NaNO3}(aq). Both the \ce{Na+} and \ce{NO3-} ions
  are, for the most part, repelled from the air-water interface ($z$ = 0). (c) Snapshot and (d) density profiles from a MD simulation
  of a 4M mixture of \ce{NaNO3} and NaBr(a) with a mole fraction of NaBr equal to 0.9. \ce{NO3-} ions are depicted as purple
  spheres, \ce{Na+} ions as green spheres, \ce{Br-} ions as orange spheres, and water molecules as blue (oxygen atoms)
  and gray (hydrogen atoms) licorice. The \ce{NO3-} density is enhanced
  below the double layer formed by adsorbed \ce{Br-} anions and \ce{Na+} counterions. The coloring of the curves in panels (c)
  and (d) correspond to the atom coloring in panels (a) and (b).}
  \label{fig:nitrate}
\end{figure}

Nitrate is one of the most abundant inorganic aerosol components in both the remote and polluted troposphere \cite{pittsbook}.
Nitrate photolysis is a source of OH \cite{mack1999,herrmann2007}, the most important oxidant in the atmosphere \cite{pittsbook}.
While the quantum yield for nitrate photolysis in bulk solution is very low ($\approx$0.06) \cite{herrmann2007}, it appears that it is
increased dramatically due to a reduced solvent cage near the air-water interface \cite{wingen2008,richards2011}.  Simulations
have predicted that in neat nitrate solutions, while the nitrate anion makes occasional excursions to the air-water interface where
it is undercoordinated, it has, on average, a relatively low intrinsic interfacial propensity (e.g., as compared to the heavier halide
anions) in large clusters and at extended aqueous surfaces \cite{minofar2006,dang2006,thomas2007,miller2009}. 
These predictions have been confirmed by SHG and PES experiments \cite{saykally2007,brown2009}. 
However, in more realistic models of sea salt aerosol that contain mixtures of sodium nitrate and sodium halide salts,
experiments have shown that the number of NO$_2$ molecules (a photolysis product) produced per NO$_3^-$ irradiated
increased with the fraction of the halide ions in the solution \cite{wingen2008,richards2011}.  MD simulations employing
polarizable force fields suggested that the electrical double layers formed by adsorbed halides followed by an excess of Na$^+$
ions draws NO$_3^-$ anions close to the solution surface where they are less solvated, and hence have reduced solvent
cages, than in neat NaNO$_3$ solutions. Figure \ref{fig:nitrate} shows how \ce{NO3-} ions are drawn closer to the air-water
interface when \ce{NaNO3} is mixed with NaBr.  This is another example of a case where the behavior of ions in mixed salt
solutions cannot be inferred from the their behavior in neat solutions.

Marine aerosols and inland aerosols originating from wind-blown dust contain a substantial amount of magnesium salts
\cite{bjfp2009pccp,gill2002}, e.g., after NaCl, the next most abundant component of seawater is magnesium chloride,
with one MgCl$_2$ for every eight NaCl \cite{crchandbook}. Mg$^{2+}$ is unusual in that it does not form ion pairs with
Cl$^-$ in aqueous solution, even at high salt concentration \cite{callahan2010}. In two recent studies, Allen, Tobias, and
co-workers used a combination of MD simulations and vibrational spectroscopy was to investigate whether or not the unique
solvation of Mg$^{2+}$ affects the distribution and solvation of Cl$^-$ near the solution-air interface of neat MgCl$_2$ over
a range of concentrations (1.1 M to 4.9 M), and a model seawater solution composed of 4.5 M NaCl and 0.3 M MgCl$_2$ 
\cite{casillas2010,callahan2010}. They found that, in spite of the fact that magnesium remains well-hydrated and has a
strong influence on the bulk structure of its solutions, its presence does not have an appreciable effect on the distribution
of chloride near the solution-air interface.  On the other hand, they noted that the presence of Mg$^{2+}$ does affect the
microsolvation of chloride in the interfacial region, and they suggested that this could, in turn, affect the heterogeneous
reactivity of chloride.

\section{SUMMARY AND OUTLOOK}
A little over a decade ago, MD simulations became established as a useful approach to gaining molecular-scale
information on chemistry occurring at aqueous interfaces in the atmosphere.  MD simulations exposed
unexpected behavior of reactive ions near the air-water interface, namely, that some ions adsorb, contrary to
longstanding conventional wisdom.  Simulation studies are helping to elucidate the driving forces for ion adsorption,
and are paving the way for new developments in theory, such as the polarizable anion dielectric continuum
theory reviewed here.  First-principles MD simulations have begun providing unprecedented details on the
solvation of ions in bulk and interfacial settings, and have started to be applied to study chemical reactions
involving ions and acids that are important in atmospheric chemistry (see, e.g.,
\cite{shamay2007,wang2009,lee2009,lewis2011,hammerich2012}).
While this review stressed the relevance of the behavior of ions at aqueous solution-air interfaces
to atmospheric chemistry, the discoveries reviewed here have led to new concepts that are much more broadly
applicable, e.g., to the broad spectrum of specific ion effects commonly referred to as Hofmeister effects.

\section*{SUMMARY POINTS}

\begin{enumerate}

\item Force field-based MD simulations and, increasingly, density functional theory-based 
MD simulations, are providing atomic-scale insight into the structure and reactivity of aqueous
interfaces in the atmosphere. 

\item In order to make progress in the validation of force fields used to
model aqueous ionic solutions, the little or no adsorption inferred from surface tension
data needs to be reconciled with the relatively strong adsorption implied by most
available spectroscopic data.

\item The polarizable anion dielectric continuum theory affords the potential of mean force for ion interactions with
the interface, from which it is possible to obtain ion distributions and, via
integration of the Gibbs adsorption isotherm, the surface tensions of various electrolyte solutions,
which are found to be in excellent agreement with experimental data.

\item Recent studies have carefully evaluated the driving forces that have 
been identified as having an important influence on the propensity for anions
to adsorb to extended aqueous interfaces. There is still considerable debate 
regarding the relative importance of induction effects, cavitation, 
ion-water and water-water interactions, local solvation effects, and the surface potential.

\item The role of the surface potential in promoting or opposing ion adsorption in point charge models of the air-water
interface is beginning to be understood, but its role in {\em ab initio} models of the interface, for which the
surface potential has the opposite sign and is an order of magnitude larger, is not yet clear.

\item The behavior of ions in concentrated mixed salt solutions cannot be inferred from the their behavior in
neat solutions. Examples include the formation of electrical double layers by adsorbed halides and their
counterions, and the enhanced interfacial population of \ce{Br-} in concentrated \ce{NaCl} solutions.  

\end{enumerate}

\section*{ACKNOWLEDGMENTS}

ACS and DJT are grateful for support from the AirUCI collaborative, which is funded by the National Science 
Foundation (grant CHE-0431312). MDB and CJM acknowledge support from the US Department of Energy's
Office of Basic Energy Sciences, Division of Chemical Sciences, Geosciences, and Biosciences.  Pacific
Northwest National Laboratory (PNNL) is operated for the Department of Energy by Battelle.  MDB is 
supported by the Linus Pauling Distinguished Postdoctoral Fellowship Program at PNNL.
YL acknowledges partial support by the CNPq, FAPERGS, INCT-FCx, and the US-AFOSR
(grant FA9550-09-1-0283).


\bibliography{arpc}{}
\bibliographystyle{unsrt}

\section*{ACRONYMS}

\begin{itemize}

\item {\bf MD:} molecular dynamics
\item {\bf GDS:} Gibbs dividing surface
\item {\bf DCT:} dielectric continuum theory
\item {\bf PA-DCT:} polarizable anion dielectric continuum theory
\item {\bf SHG:} second harmonic generation
\item {\bf VSFG:} vibrational sum frequency generation
\item {\bf PES:} photoelectron spectroscopy
\item {\bf OS:} Onsager-Samaras
\item {\bf DFT:} density functional theory
\item {\bf EXAFS:} extended X-ray absorption fine structure
\item {\bf PMF:} potential of mean force

\end{itemize}

\section*{KEY TERMS}

\begin{itemize}

\item {\bf Aerosol:} stable suspension of particles in the atmosphere

\item {\bf Troposphere:} region of the atmosphere between ground level and about 15 km altitude

\item {\bf Gibbs dividing surface:} position in interface where the surface excess of solvent is zero

\item {\bf Poisson-Boltzmann theory:} mean field theory for ion distributions derived by combining
Poisson's equation with the Boltzmann distribution

\item {\bf Gibbs adsorption isotherm:} equation relating surface excess of a solute to derivative
of the surface tension with respect to solute concentration

\item {\bf Generalized Gradient Approximation:} class of exchange-correlation functionals that
includes the electron density and its gradient

\item {\bf Hofmeister series:} ordered ranking of ions according to their influence on a wide variety 
of chemical and biological processes. 

\item {\bf QM/MM:} hybrid simulation technique in which one region is treated quantum mechanically
while the rest of the system is modeled with molecular mechanics.

\item{\bf Surface potential:} electric potential difference across an interface that arises from broken symmetry

\end{itemize}

\section*{ANNOTATED REFERENCES}

\begin{itemize}

\item {\bf A. Arslanargin and T. L. Beck. Free energy partitioning analysis of the
driving forces that determine ion density profiles near the water liquid-vapor
interface. {\em J. Chem. Phys.}, 136:104503, 2012.}

Decomposition of ion adsorption free energy into contributions 
from cavity formation, dispersion, and local and far-field electrostatics.

\item {\bf M. D. Baer and C. J. Mundy. Towards an understanding of the specific ion
effect using density functional theory. {\em J. Phys. Chem. Lett.}, 2:1088--1093, 2011.}

PMF of transferring an iodide from bulk to interface is computed using
density functional theory.

\item {\bf C. Caleman, J. S. Hub, P. J. van Maaren, and D. van der Spoel. Atomistic 
simulation of ion solvation in water explains surface preference of halides. {\em Proc. 
Natl. Acad. Sci. USA}, 108:6838--6842, 2011.}

Reports single halide ion PMFs and their decomposition into enthalpic and entropic contributions.

\item {\bf S. Ghosal, J. C. Hemminger, H. Bluhm, B. S. Mun, E. L. D. Hebenstreit,
G. Ketteler, D. F. Ogletree, F. G. Requejo, and M. Salmeron. Electron
spectroscopy of aqueous solution interfaces reveals surface enhancement of
halides. {\em Science}, 307:563--566, 2005.}

Composition of the liquid/vapor interface is measured using
x-ray photoelectron spectroscopy.

\item {\bf P. Jungwirth and D. J. Tobias. Molecular structure of salt solutions: A
new view of the interface with implications for heterogeneous atmospheric
chemistry. {\em J. Phys. Chem. B}, 105:10468--10472, 2001.}

Pioneering molecular dynamics simulation study of ion distributions 
at extended sodium halide solution-air interfaces.



\item {\bf Y. Levin. Polarizable ions at interfaces. {\em Phys. Rev. Lett.}, 102:147803, 2009.}

Presents a polarizable anion dielectric continuum theory 
that agrees well with experimental surface tension and surface potential data.

\item {\bf D. Liu, G. Ma, L. M. Levering, and H. C. Allen. Vibrational spectroscopy
of aqueous sodium halide solutions and air-liquid interfaces: Observation of
increased interfacial depth. {\em J. Phys. Chem. B}, 108:2252--2260, 2004.}

Perturbation of interfacial water hydrogen-bonding by bromide and 
iodide anions revealed for the first time by vibrational sum frequency generation.

\item {\bf D. E. Otten, P. R. Shaffer, P. L. Geissler, and R. J. Saykally. Elucidating
the mechanism of selective ion adsorption to the liquid water surface. {\em Proc.
Natl. Acad. Sci. USA}, 109:701--705, 2012.}

UV second harmonic generation spectroscopy and simulations demonstrate enthalpic promotion and
entropic opposition to anion adsorption.

\item {\bf P. B. Petersen and R. J. Saykally. Confirmation of enhanced anion 
concentration at the liquid water surface. {\em Chem. Phys. Lett.}, 397:51--55, 2004.}

First verification of enhanced concentration of anions at the solution interface by
second harmonic generation.

\item {\bf J. Schmidt, J. Vandevondele, I.-F. W. Kuo, D. Sebastiani, J. I. Siepmann,
J. Hutter, and C. J. Mundy. Isobaric-isothermal molecular dynamics simula-
tions utilizing density functional theory: An assessment of the structure and
density of water at near-ambient conditions. {\em J. Phys. Chem. B}, 113:11959--11964, 2009.}

Shows that the Grimme empirical dispersion correction to DFT improves water density
and structure in constant pressure MD simulations.  

\end{itemize}

\end{document}